\documentclass{article}
\usepackage{spconf,amsmath,graphicx}

\title{NN3A: Neural Network supported Acoustic Echo Cancellation, Noise Suppression and Automatic Gain Control for Real-Time Communications}
\name{Ziteng Wang, Yueyue Na, Biao Tian, Qiang Fu}
\address{Alibaba Group, China}
\begin{document}
%
\maketitle
\begin{abstract}
Acoustic echo cancellation (AEC), noise suppression (NS) and automatic gain control (AGC) are three often required modules for real-time communications (RTC). This paper proposes a neural network supported algorithm for RTC, namely NN3A, which incorporates an adaptive filter and a multi-task model for residual echo suppression, noise reduction and near-end speech activity detection.
The proposed algorithm is shown to outperform both a method using separate models and an end-to-end alternative.
It is further shown that there exists a trade-off in the model between residual suppression and near-end speech distortion, which could be balanced by a novel loss weighting function.
Several practical aspects of training the joint model are also investigated to push its performance to limit.
\end{abstract}
\section{Introduction}

\begin{figure}[htb]
	\centering
	\centerline{\includegraphics[width=\columnwidth]{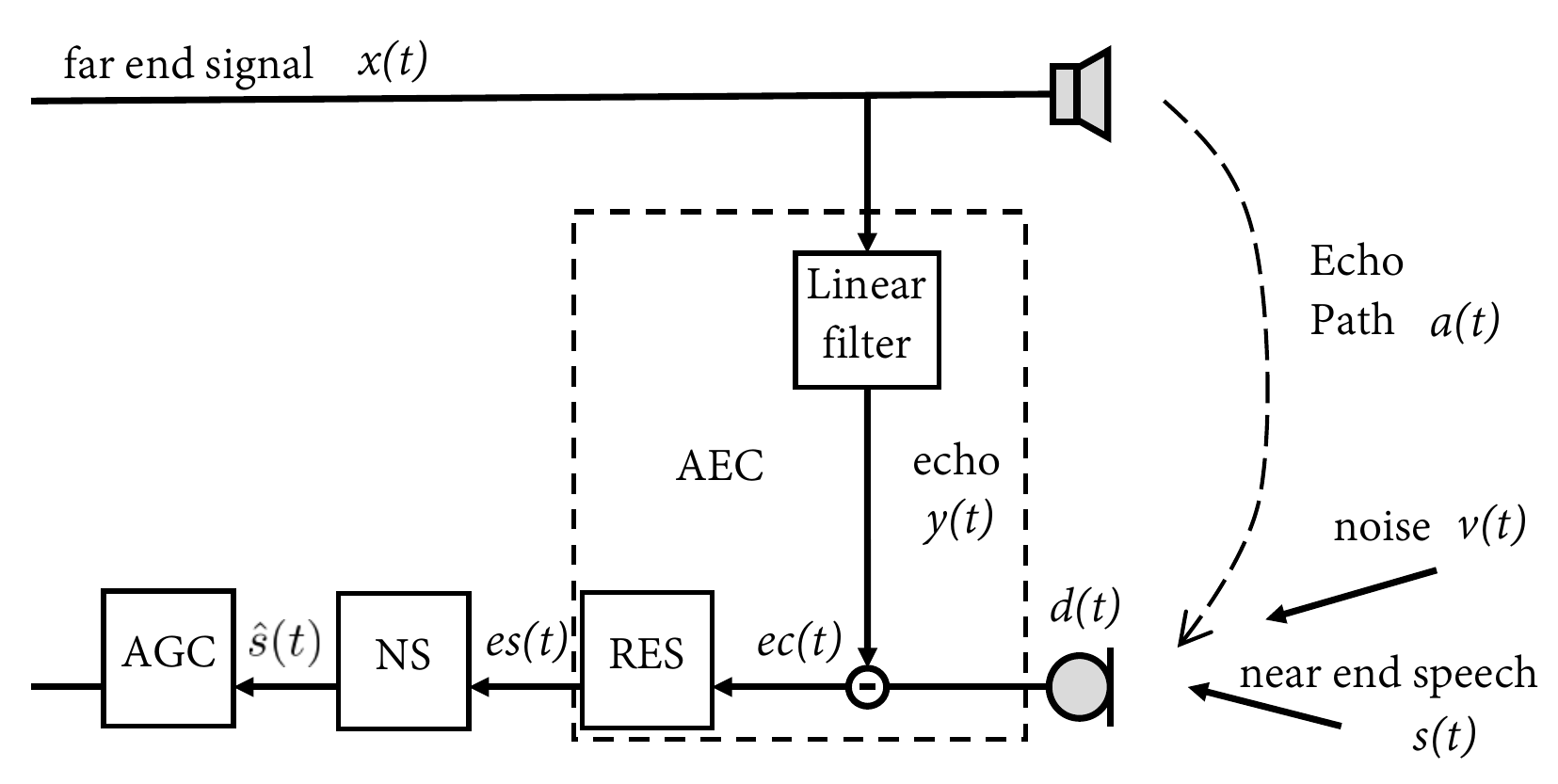}}
	\caption{Typical AEC/NS/AGC modules for real-time communications.}
	\label{fig:rtc}
\end{figure}

The demand for real-time communications (RTC) has grown rapidly in recent years. The three often required modules for RTC are acoustic echo cancellation (AEC), noise suppression (NS) and automatic gain control (AGC), as shown in Fig.\ref{fig:rtc}. AEC is designed to remove echo from the near-end microphone signal. It usually consists of a linear echo cancellation filter and a residual echo suppressor (RES). NS is subsequently to reduce the background noise, and AGC is to adjust the processed signal to a proper sound level.
Many neural network based algorithms have been developed to tackle each problem separately, and casual and real-time ones are preferred as seen in the recent DNS-Challenges~\cite{reddy2021icassp,reddy2021interspeech} and AEC-Challenges~\cite{sridhar2021icassp,cutler2021interspeech}.

The deep learning based algorithms first made breakthrough in the NS task. Xu et al. \cite{xu2013experimental} proposed to find a spectral mapping function between noisy and clean speech signals. Wang et al. \cite{wang2014training} proposed to learn a mapping from noisy features to time-frequency masks of the target, such as the ideal binary mask, the phase-sensitive mask (PSM)~\cite{erdogan2015phase}, or the later introduced complex ratio mask (CRM)~\cite{williamson2015complex}. 
Advanced neural networks have been developed, such as the convolutional time-domain audio separation network (Conv-TasNet)~\cite{luo2019conv} and the deep complex convolution recurrent network (DCCRN)~\cite{hu2020dccrn} , and remarkable results are achieved.

The neural network based AEC algorithms follow a similar paradigm by considering the echo signal as \emph{undesired noise}.
\cite{zhang2019deep,westhausen2021acoustic} investigated end-to-end AEC relying on only neural networks.
Nevertheless, a joint approach of linear filtering and neural network based RES has shown to be more effective~\cite{wang2021weighted,valin2021low}. 
In the latter case, the linear filtered output, the microphone signal, the far-end signal, as well as the estimated echo, could all be taken as model  inputs~\cite{carbajal2018multiple,fazel2020cad,franzen2021deep}.

The need for an AGC is to amplify speech segments to an intelligible sound level, while not amplifying noise only segments~\cite{chu1996voice,archibald2008software}. This can be properly handled once the near-end speech activity results are obtained, usually from a model based voice activity detector (VAD)~\cite{hughes2013recurrent,braun2021training}.

Though the three modules each involves a neural network model, they are separately designed and a joint model remains not investigated.
The motivations are: a neural network model trained to suppress echo is likely to reduce noise at the same time, and multi-task training could benefit the overall performance.
A joint model is also more compact from the engineering perspective.
In this paper, we consider the three modules systematically and propose a NN3A algorithm, building upon our previous work~\cite{wang2021weighted} for echo cancellation. The NN3A algorithm retains a linear adaptive filter and adopts a multi-task model for joint residual echo suppression, noise reduction and near-end speech activity detection.
It is shown that there exists a trade-off in the model between residual suppression and near-end speech distortions. And an intuitive loss weighting function is introduced to balance the trade-off to meet the \emph{zero} echo leakage requirement in typical RTC applications.
The proposed algorithm outperforms both a method using separate models and an end-to-end neural network based alternative.


\section{The NN3A algorithm}

Consider a typical single microphone single loudspeaker setup, the microphone signal at time $t$ is expressed as:
\begin{equation}\label{eq:mic}
d(t) = x(t)*a(t) + s(t) + v(t)
\end{equation}
where $x(t), s(t)$ and $v(t)$ are respectively the far-end signal, the near-end speech and the ambient noise. $a(t)$ is the echo path and $*$ denotes convolution. 
The task of AEC, NS and AGC is to recover a properly scaled version of the near-end speech $s(t)$, given the microphone signal and the far-end signal.

Without loss of generality, the algorithm is developed in the frequency domain.
The frequency representations of the variables are denoted by their capitals, such as $D, X, S$.

\subsection{Linear filter}

A linear adaptive filter is first adopted to remove an echo estimation $Y$ from the microphone signal:
\begin{equation}
E_{t,f} = D_{t,f} - \underset{Y_{t,f}}{\underbrace{ {\bf w}^H_{L, f} {\bf x}_{L, f} }}
\end{equation}
where ${\bf x}_{L,f} = [X_{t,f}, X_{t-1,f}, ..., X_{t-L+1,f}]^T$, $f$ is the frequency index, $(\cdot)^T$ denotes transpose and $(\cdot)^H$ denotes Hermitian transpose. $L$ is the filter tap and the filter coefficients are derived using the weighted recursive least square (wRLS) algorithm \cite{wang2021weighted} as:
\begin{align}
\nonumber \gamma_{t,f} &\leftarrow |E_{t,f}|^{\beta - 2}, \\
\nonumber {\bf R}_{L,f} &\leftarrow \sum_{t} \gamma_{t,f}{\bf x}_{L,f}{\bf x}_{L,f}^H, \\
\nonumber {\bf r}_{L,f} &\leftarrow \sum_{t} \gamma_{t,f}{\bf x}_{L,f}D_{t,f}^*,\\
{\bf w}_{L, f} &\leftarrow {\bf R}_{L, f}^{-1}{\bf r}_{L,f}
\end{align}
where $\beta \in [0, 2]$ is a shape parameter related to the speech source prior, and $(\cdot)^*$ denotes complex conjugate.

\subsection{Multi-task model}

After linear filtering, a neural network based multi-task model is designed to further suppress the residual echo and remove the ambient noise:
\begin{equation}
	\hat{S}_{t,f} = M_{t,f} E_{t,f}
\end{equation}
where $M_{t,f}$ is the time-frequency mask inferred from the available signal set ${\bf f}_{t} = \{E_{t,f}, Y_{t,f}, D_{t,f}, X_{t,f} \}$. The inference process is expressed as:
\begin{align}
\label{eq:nnres}
\nonumber {\bf h}^0_t & = \text{ReLU}(\text{Linear}({\bf f}_{t})) \\
\nonumber {\bf h}^{j+1}_t & = \text{DFSMN}({\bf h}^{j}_t), \quad j\in[0,1,...,J-1] \\
\nonumber M_{t,-} & = \text{Sigmoid}(\text{Linear}({\bf h}^{j+1}_t)) \\
P_t & = \text{Sigmoid}(\text{Linear}({\bf h}^{j+1}_t))
\end{align}
where ${\bf h}^{j}_t$ is output of the $j$th layer, $P_t$ is the probability of near-end speech activity, and Linear($\cdot$) denotes an affine transformation layer with proper size of parameters. The deep feed-forward sequential memory network (DFSMN)~\cite{zhang2018deep} is chosen to model temporal dependency in time series, and one DFSMN layer is expressed as: 
\begin{align}
\nonumber \tilde{{\bf h}}^{j}_{t} &= \text{Linear}(\text{ReLU}(\text{Linear}({\bf h}^{j-1}_t))), \\
{\bf h}^{j}_{t} &= {\bf h}^{j-1}_{t} + \tilde{{\bf h}}^{j}_{t} + \sum_{\tau=0}^{\mathcal{T}}{\bf m}^{j}_{\tau} \odot \tilde{{\bf h}}^{j}_{t-\tau}
\end{align}
where ${\bf m}_{\tau}$ is the time-invariant memory parameter weighting the history output $\tilde{{\bf h}}_{t-\tau}$ and $\odot$ denotes element-wise multiplication.

Given a predefined training target, such as the PSM
\begin{equation}
	 \bar{M}_{t,f}=\text{Real}(\frac{S_{t,f}}{E_{t,f}}),
\end{equation}
the network model is trained to minimize both a mean squared error loss $\mathcal{L}_{mask}$ and a binary cross-entropy loss $\mathcal{L}_{vad}$, where
\begin{align}
\nonumber	\mathcal{L}_{mask} &= \sum_{t,f} \alpha_{t,f} |M_{t,f} - {\bar{M}}_{t,f}|^2, \\
	\mathcal{L}_{vad} &= \sum_{t} -\bar{P}_t \log(P_t) - (1-\bar{P}_t) \log(1-P_t)
\end{align}
where $\bar{P}_t \in \{0,1\}$ is the oracle near-end speech activity. 
It is empirically found that a model trained under the vanilla MSE loss cannot completely remove the residual echo, which fails to meet the usual human auditory requirement of zero echo leakage. Hence a weighting function is introduced as:
\begin{equation}\label{eq:lw}
	\alpha_{t,f} = \alpha - \bar{M}_{t,f}, \quad \alpha > 1
\end{equation}
The weighting function puts more weight on echo dominant time-frequency bins, where $\bar{M}_{t,f}$ is small, and less weight otherwise.

\begin{figure*}[t]
	\centering
	\includegraphics[width=15cm]{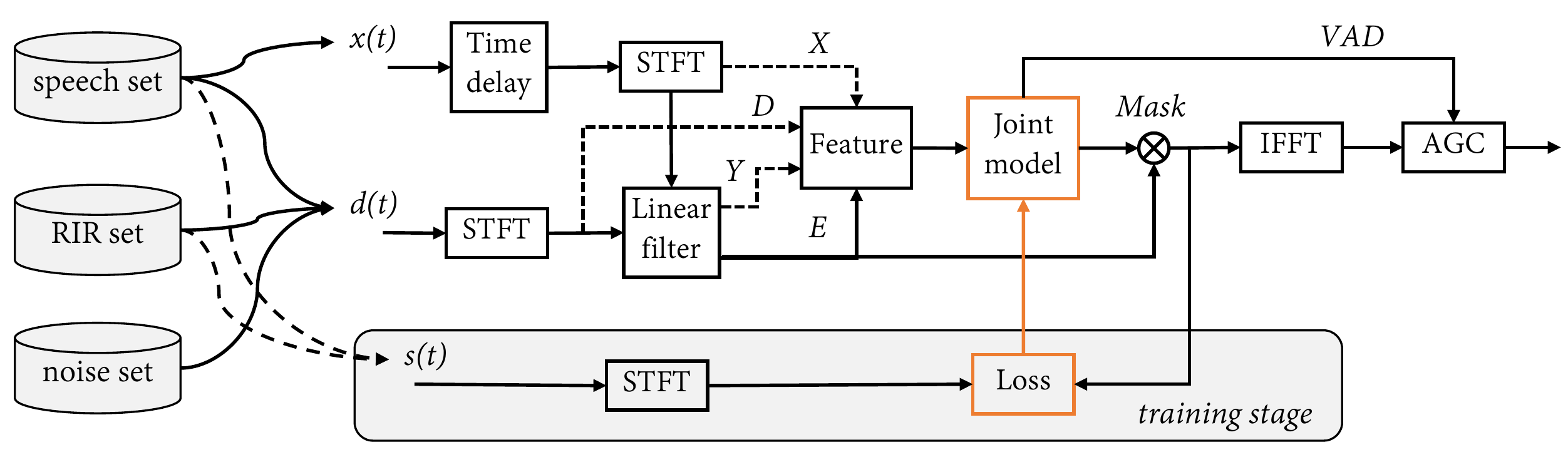}
	\caption{Flowchart of the proposed NN3A algorithm. The gray blocks are only for the training stage.}
	\label{fig2}
\end{figure*}

\subsection{Post-processing}

The sound level of the processed signal is subject to change in different applications. Therefore, the AGC algorithm is put in a post-processing stage as
\begin{equation}
\hat{s}(t) = g(P_t)\text{IFFT}(\hat{S}_{t,f})
\end{equation}
and $g(\cdot)$ is a customized function, which consists of a peak level detector for computing gain and a gain controller for adjusting gain\cite{archibald2008software}.

The flowchart of the proposed NN3A algorithm is presented in Fig~\ref{fig2}. To handle possible delays between the microphone signal and the far-end signal, an additional time delay compensation module could be added before processing~\cite{wang2021weighted}.

\subsection{Related work}

Combining speech activity detection and echo suppression has been discussed in \cite{zhang2019deep}. Nevertheless, the study considered a pure model based approach and the experiments were limited to small-scale simulated data.
An intuitive loss weighting method was proposed in~\cite{9414958} for spectral mapping based echo suppression, while a risk of sub-band nullification was introduced at the same time.

\section{Experiments}

\subsection{Setup}

Experiments are conducted following the setup in the AEC-Challenge~\cite{sridhar2021icassp}. 
The training set covers near-end (NE) single talk (ST), far-end (FE) single talk, and double talk (DT) cases.
Besides the data provided in the Challenge, a simulation pipeline is built to generate more training samples as in Fig~\ref{fig2}.
The speech set, the room impulse response (RIR) set and the noise set are all drawn from the DNS-Challenge~\cite{reddy2021icassp}.
The signal-to-echo ratio (SER) are uniformly sampled from [-10, 20] dB, and the signal-to-noise (SNR) ratio are sampled from [0, 40] dB.
The simulated data are mixed as in (\ref{eq:mic}), with 30\% $x(t)=0$, 20\% $v(t)=0$, and 10\% $a(t)=0$ implying a muted loudspeaker.
Finally, approximate 1k hours of training data are used during training.

\begin{table}[th]
	\begin{center}
		\caption{Step-wise evaluation of the proposed algorithm. + denotes cascade, and \& denotes joint processing.}
		\label{tab1}
		\begin{tabular}{l|c|c|c}
			\hline 
			& \begin{tabular}{@{}c@{}}NE ST \\ PESQ\end{tabular} 
			& \begin{tabular}{@{}c@{}}FE ST \\ ERLE\end{tabular} 
			& \begin{tabular}{@{}c@{}}DT \\ PESQ \end{tabular} \\ \hline
			Orig &1.65&	0	   &1.86  \\ \hline
			Linear &1.65& ~~5.49 	&2.23  \\ \hline
			+RES &1.84&	34.39	&2.70  \\ \hline
			+RES+NS &2.49&	38.28	&2.72 \\ \hline
			+RES\&NS &2.42&	35.11	&2.75  \\ \hline
			NN3A, $\alpha_{t,f}=1$ &2.47	&37.25 &\bf 2.79 \\ \hline
			NN3A, $\alpha=1.1$&\bf 2.57&	\bf 45.13	&2.69 \\ \hline
		\end{tabular}
	\end{center}
\end{table}

\begin{table}[th]
	\begin{center}
		\caption{Comparison of different multi-task model inputs ${\bf f}_{t} = \{E, Y, D, X \}$, as well as two end-to-end setups.}
		\label{tab2}
		\begin{tabular}{l|c|c|c}
			\hline 
			& \begin{tabular}{@{}c@{}}NE ST \\ PESQ\end{tabular} 
			& \begin{tabular}{@{}c@{}}FE ST \\ ERLE\end{tabular} 
			& \begin{tabular}{@{}c@{}}DT \\ PESQ \end{tabular} \\ \hline
			EX &2.57&	45.13&	2.69  \\ \hline
			EY &2.58&	52.79&	2.71  \\ \hline
			EYD &\bf 2.61&	53.37&	\bf 2.75  \\ \hline
			EYDX &2.59&	\bf 53.99&	2.74 \\ \hline \hline
			DX &2.54&	43.59&	2.52  \\ \hline
			DTLN~\cite{westhausen2021acoustic} &2.25&	34.01&	2.65 \\ \hline
		\end{tabular}
	\end{center}
\end{table}

For STFT, the frame size is 20 ms and the hop size is 10ms. For the wRLS filter, the filter tap $L=5$, and the source prior shape parameter $\beta=0.2$. The multi-task model consists of $J=12$ DFSMN layers, each layer with 512 nodes, and the temporal order $\mathcal{T} = 20$, leading to 6.7 M parameters.

\subsection{Evaluation set}

The evaluation set consists of three parts:

\textbf{NE ST}: 480 near-end speech utterances, contaminated by noise types of \{car, subway, howling, keyboard, office, white\} with SNR of \{-5, 0, 5, 10\} dB.

\textbf{FE ST}: 300 far-end single talk real recordings from the Challenge blind testset.

\textbf{DT}: the top 500 double talk utterances in the Challenge synthetic set.

The ITU-T recommendation P.862 perceptual evaluation of speech quality (PESQ) and echo return loss enhancement (ERLE) scores are reported.

\subsection{Results and Analysis}

We first show step-wise evaluations of the algorithm in Table~\ref{tab1}. A model that only performs residual echo suppression is denoted as RES, which also shows the ability to suppress noise, increasing the PESQ score by 0.19 for noisy NE ST.
Stacking RES with a separately trained NS model, the RES+NS setup outperforms a joint RES\&NS model in suppressing residuals. But cascading two models is likely to over-suppress the near-end speech, as shown by the lower DT PESQ score (2.72 compared with 2.75).
The proposed vanilla NN3A algorithm (with $\alpha_{t,f}=1$), which combines residual suppression with near-end speech activity detection, improves the results overall.

\begin{figure}[t]
	\centering
	\includegraphics[width=\columnwidth]{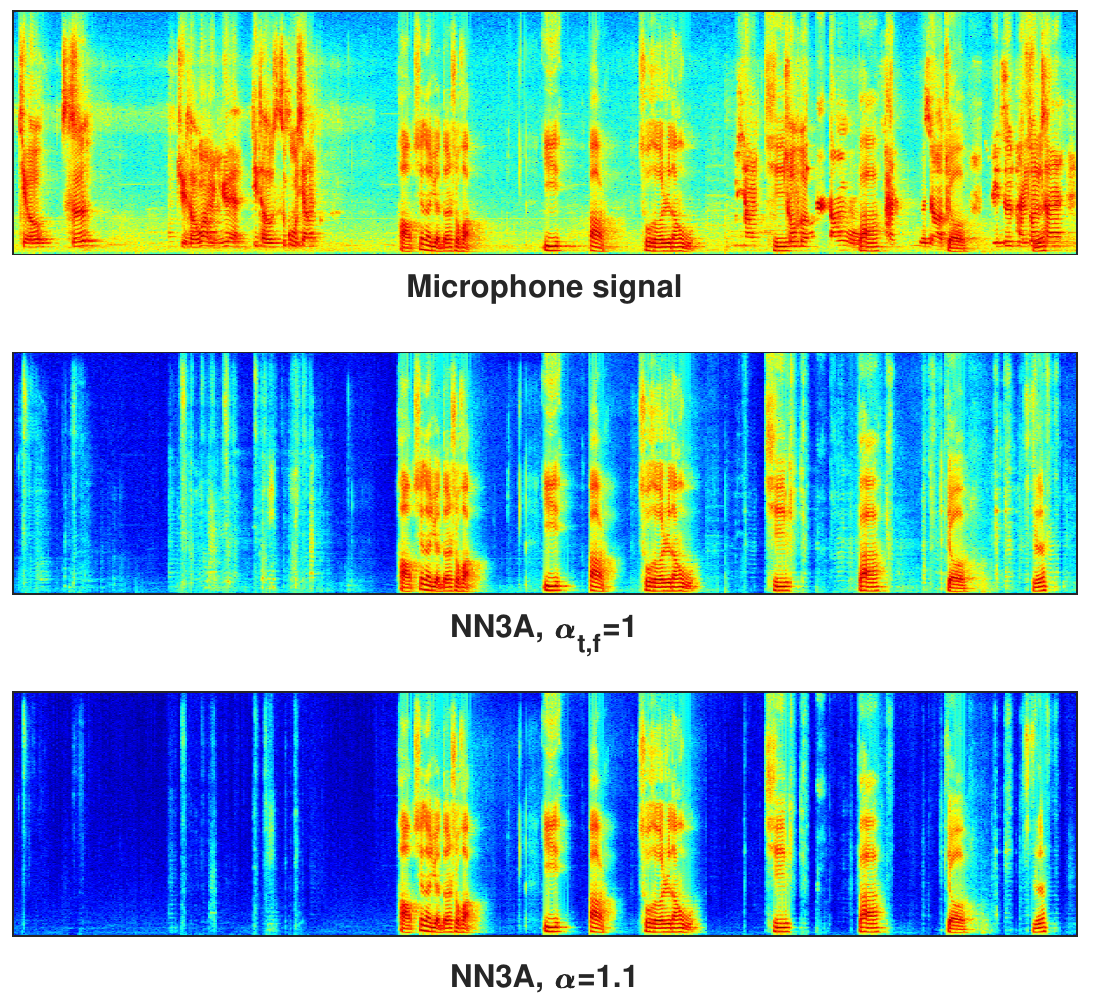}
	\caption{Illustration of the effect of loss weighting. The sample audio covers FE ST, NE ST and DT consecutively.}
	\label{fig-lw}
\end{figure}

To achieve minimum echo leakage, $\alpha=1.1$ is set in (\ref{eq:lw}). As expected, the residual suppression performance is largely improved with a ERLE score of 45.13 dB compared with the 37.25 dB reference, at the cost of more speech distortion. An audio sample is given in Fig.~\ref{fig-lw} to illustrate the effect of loss weighting. The residual echo in the last signal is finally reduced to below audible level.

In Table~\ref{tab2}, possible algorithm configurations are investigated to improve the performance. It is observed that a combination of the linear filtered output $E$, the estimated echo $Y$ and the microphone signal $D$, leads to much higher scores than our previous EX setup~\cite{wang2021weighted}. And using the estimated echo as input is a better choice than using the far-end signal.
The findings are consistent with a recent preprint~\cite{franzen2021deep}.

The NN3A algorithm is also compared with a setup that discards the linear filter, denoted as DX, and a publicly available DTLN (10.4 M) model~\cite{westhausen2021acoustic}.
There is a clear advantage of the joint signal processing and deep learning approach over the pure model based ones.
In Fig.~\ref{fig-lw}, the DX setup is compared with the NN3A-EYD setup in different SERs.
The results verifies the benefit of linear filtering, and the gap is larger especially in low SER scenarios.

\begin{figure}[t]
	\centering
	\includegraphics[width=\columnwidth]{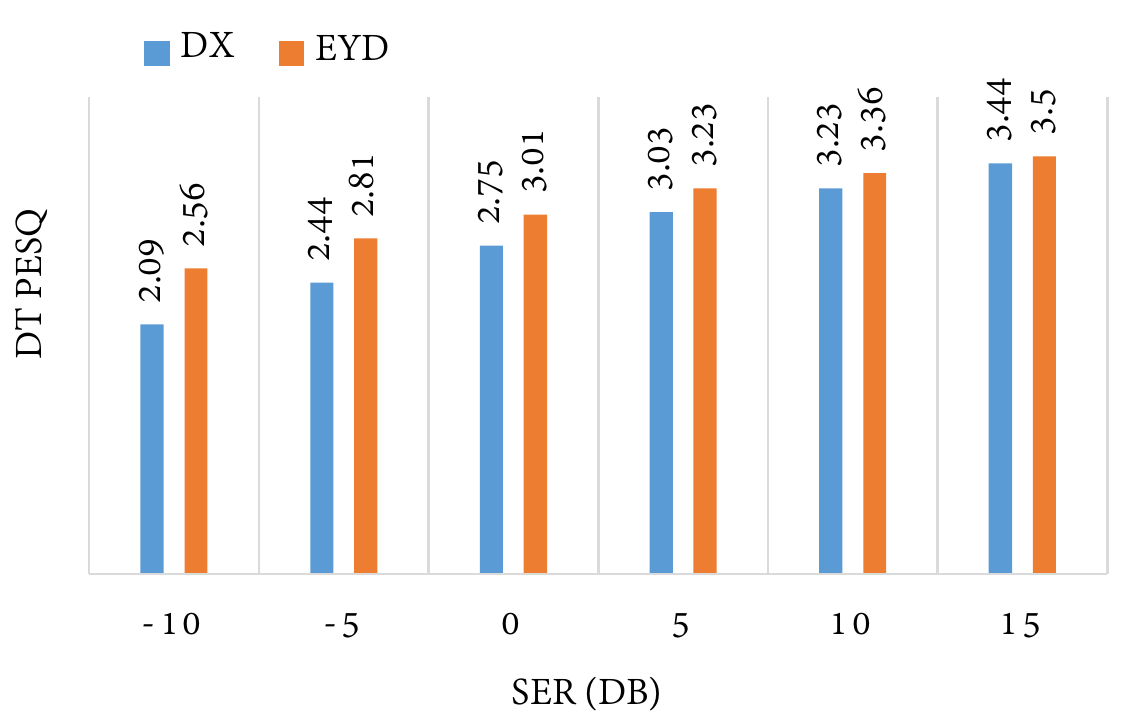}
	\caption{Comparison of an end-to-end setup (DX) and the NN3A algorithm (EYD) in different SERs.}
	\label{fig3}
\end{figure}

\section{Conclusion}

This paper presents a NN3A algorithm for real-time communications. The algorithm retains a linear filter and introduces a multi-task model for joint residual echo suppression, noise reduction and near-end speech activity detection. 
The multi-task model is shown to outperform both cascaded models and end-to-end alternatives. 
A novel loss weighting function is introduced to balance tradeoff between residual suppression and speech distortion, and minimum echo leakage could be achieved by tuning a weighting factor.

\bibliographystyle{IEEEbib}
\bibliography{refs}

\end{document}